# Fe XVII X-ray Line Ratios for Accurate Astrophysical Plasma Diagnostics

J. D. Gillaspy[1], T. Lin[2], L. Tedesco[1*], J. N. Tan[1], J. M. Pomeroy[1], J. M. Laming[3], N. Brickhouse[2], G.-X. Chen[2], and E. Silver[2]

[1]National Institute of Standards and Technology, Gaithersburg, MD 20899-8422, USA
[2]Harvard-Smithsonian Center for Astrophysics, 60 Garden Street, Cambridge, MA 02138, USA
[3]Space Science Division, Naval Research Laboratory, Code 7674L, Washington, DC 20375, USA

email address of corresponding author: john.gillaspy@nist.gov

---

[*] current address: Brown University, Providence, RI 02912

# ABSTRACT

New laboratory measurements using an Electron Beam Ion Trap (EBIT) and an x-ray microcalorimeter are presented for the n=3 to n=2 Fe XVII emission lines in the 15 Å to 17 Å range, along with new theoretical predictions for a variety of electron energy distributions. This work improves upon our earlier work on these lines by providing measurements at more electron impact energies (seven values from 846 to 1185 eV), performing an in situ determination of the x-ray window transmission, taking steps to minimize the ion impurity concentrations, correcting the electron energies for space charge shifts, and estimating the residual electron energy uncertainties. The results for the *3C/3D* and *3s/3C* line ratios are generally in agreement with the closest theory to within 10%, and in agreement with previous measurements from an independent group to within 20%. Better consistency between the two experimental groups is obtained at the lowest electron energies by using theory to interpolate, taking into account the significantly different electron energy distributions. Evidence for resonance collision effects in the spectra is discussed. Renormalized values for the absolute cross sections of the *3C* and *3D* lines are obtained by combining previously published results, and shown to be in agreement with the predictions of converged R-matrix theory. This work establishes consistency between results from independent laboratories and improves the reliability of these lines for astrophysical diagnostics. Factors that should be taken into account for accurate diagnostics are discussed, including electron energy distribution, polarization, absorption/scattering, and line blends.

1. INTRODUCTION

The x-ray lines in Ne-like Fe XVII in the 700 eV to 950 eV (18 Å to 13 Å) spectral range are among the strongest and most ubiquitous lines observed by Chandra and XMM Newton, the leading x-ray spectroscopy observatories of the past decade. Because the intensity ratios of these lines are sensitive to a variety of local plasma conditions, they are prime candidates for astrophysical diagnostics (Audard et al. 2004; Drake et al. 1999; Gudel 2004; Huenemoerder, Testa, & Buzasi 2006; Mewe et al. 2001; Osten et al. 2003). Yet their use has been hampered by inconsistencies between (a) various laboratory experiments, when compared to each other, (b) various predictions of theory, when compared to each other, (c) the predictions of theory when compared to the laboratory experiments, and (d) astrophysical conclusions drawn using these lines in comparison to the conclusions drawn using other lines. The lines are of such potential value for astrophysical diagnostics, however, that the above problems have stimulated a wide range of experimental and theoretical activity from a variety of groups, including (Aggarwal, Keenan, & Msezane 2003; Beiersdorfer et al. 2002; Bhatia & Saba 2001; Brown et al. 2001a; Brown et al. 1998; Brown, Beiersdorfer, & Widmann 2001b; Chen et al. 2006; Doron & Behar 2002; Fournier & Hansen 2005; Laming et al. 2000; Loch et al. 2006), which have aimed at establishing consistency. In this paper, we present new data that successfully establish consistency between theory and the most accurate measurements from two independent laboratories. We also discuss the extent to which resonance enhancement of the collisions can be seen in our data.

The lines under consideration are listed in table I and shown in figure 1. We refer to these lines throughout this paper using the capital letters in the second column, following the conventional notation of Parkinson (1973), but we drop the prefix 3 that is common to all lines under consideration here (e.g., we refer to 3C simply as C). We refer to the commonly discussed *groups* of lines using the lower case letters "s" and "d" as indicated in the table. In this paper, we focus on the line ratios *C/D* and *s/C*, as these are the ones that have been most extensively discussed in the laboratory astrophysics literature recently. The wavelengths in table I are from (Brown, et al. 1998; Shirai et al. 2000).

For the C/D ratio, previous measurements from the Lawrence Livermore National Laboratory (LLNL) electron beam ion trap (EBIT) (Brown, et al. 1998) and the NIST EBIT (Laming, et al. 2000) have been in relatively good agreement with each other and with theory (Chen & Pradhan 2002). Although the value reported from the NIST EBIT at 900 eV (Laming, et al. 2000) was 6% higher than the value reported by the LLNL group at 850 eV (Brown, et al. 1998), the difference was within the 6%-7% experimental uncertainties and had the same sign as the predicted energy-dependence. The NIST value at 1250 eV (Laming, et al. 2000) was 15% lower than the lowest of two values measured by LLNL at 1150 eV (Brown, et al. 1998), but in this case the error bars nearly touch and the discrepancy is of marginal statistical significance.

Subsequently, the LLNL group showed (Brown, et al. 2001a) that when Fe was loaded into their trap as a gas, they could vary their measured C/D ratio far outside of their original error bars by raising the gas pressure to high values. They attributed this to a modification of the charge balance and the existence of overlapping lines in Fe XVII and other nearby charge states of Fe. They found that when they injected ions using a metal vapor vacuum arc (the method we use in all of our work on metals), this problem was absent. Our subsequent work on Ne-like Ni (slightly higher up the isoelectronic sequence) showed good agreement with the predicted energy dependence of the *C/D* ratio (Chen, et al. 2006) and with previous measurements by the LLNL group (Brown, et al. 2001b; Gu et al. 2004). In fact, these benchmarked results have been used to revisit the issue of the transparency of the solar corona (Brickhouse & Schmelz 2006) with the result reversing earlier conclusions and leading to consequences relevant to the planning of future space missions. For example, it was shown that the fuzziness of certain images is a result of a high density of coronal structures, rather than scattering, and hence observations at higher spatial resolution could resolve loop structure at temperatures closer to those at which the dominant heating occurs, thus advancing efforts to understand coronal heating.

Despite the generally satisfactory results for the ratio discussed above, it has been suggested (Brown et al. 2006) that the calculated absolute values of both the *C* and *D* cross sections contain relatively large errors (25% to 50%) that approximately cancel in the ratio. As we show in section 5 below, however, this conclusion does not hold if one uses the most recent set of radiative recombination (RR) and electron impact excitation (EIE) cross section calculations made by one of us (Chen 2007, 2008a).

The situation with regard to the more widely separated *s* and *d* groups of lines, however, has up to now remained problematic. Discrepancies between laboratory measurements for the s/d ratio have been reported to be as large as 50% (Beiersdorfer, et al. 2002). We discuss in sections 3.2 and 3.3 below several reasons why the discrepancy was actually closer to 30%, but the general conclusion remains that prior to the present work, the discrepancy between laboratory measurements was larger than the precision needed for modern astronomy. Uncertainties of 25% to 30% in line ratios have been shown to be insufficient for astrophysical diagnostics using Ne IX, for example (Smith et al. 2009). This need for higher accuracy forms the motivation for the present work, in which we record a fresh set of data under conditions relatively immune to several possible sources of systematic error in the previous results. We then put particular effort into adjusting for all remaining systematic errors and accurately estimating both the horizontal (energy) and vertical (line ratio) uncertainties, the latter of which now range from 3% to 10%. We also follow (Brown, et al. 2001a) in using *s/C* (rather than *s/d* ) as a more robust line ratio which is less sensitive to line blends. We note, however, that potential line blend problems are not entirely removed in this ratio unless the resolution is << 6 eV, a condition not typically fulfilled by the microcalorimeters used on EBITs to date.

In our original publication (Laming, et al. 2000), we noted concerns about the uncertainty in the window transmission function, which can alter the observed line ratios. Our results in that paper hinged on calculated values based on window thicknesses measured independently long before that experiment began. We pointed out that it was possible that the transmission function had changed due to the buildup of minute amounts of hydrocarbons or cryo-condensed gases, for example. The relatively wide separation of photon energies in the *s/C* ratio makes assigning experimental uncertainties due to the variation in window transmission more difficult than in the case of the *C/D* ratio. The present work thus includes a system designed to quantify and monitor the window transmission function during the course of the experiment. We found only a small difference between our estimated and measured window transmission function in the present work (corresponding to a shift in the *s/C* ratio of approximately 7%), although we cannot rule out the possibility that a more significant difference was present at the time of our previous work.

Other improvements in our experiment discussed below include a more detailed determination of the electron beam energy and attention to the purity of the charge state and species in the trap. At the time of our first experiments, neither of these factors were expected to influence the results substantially, but subsequent work has quantified the extent to which they can indeed be significant (Chen 2007; Chen, et al. 2006; Chen & Pradhan 2002; Brown 2001a). We use only a microcalorimeter, which is insensitive to polarization and therefore does not require large systematic adjustments to the measured data.

## 2. EXPERIMENT AND RAW DATA

The highly charged ions were produced and held in the NIST Electron Beam Ion Trap (Gillaspy 1997) at beam energies of 846(18) eV, 857(13) eV, 902(19) eV, 947(21) eV, 982(22) eV, 1094(22) eV, and 1185(22) eV, and beam currents of 16 mA, 12 mA, 18 mA, 20 mA, 21 mA, 23 mA, and 23.5 mA, respectively. These values for the energies include the effects of space charge, and are lower than the nominal values determined by the potentials applied to the drift tubes, as described below. The beam was tuned, as described elsewhere (Gillaspy 1997), to keep the reflected current less than 20 μA (< 0.2% of the beam current). The magnetic induction at the trap center was 2.7 T and the electrostatic trap depth was 220 V.

In order to optimize the purity of the charge state distribution and avoid potential problems discussed previously (Brown, et al. 2001a), several measures were taken. The trap was emptied and reloaded with Fe ions every 2.2 seconds using a metal vapor vacuum arc (Holland et al. 2005). Furthermore, the trap was warmed up from its operating temperature of 4.2K to ≥77K and pumped out for several weeks prior to the present experiment. During the experiment, the only evaporative cooling gas used was that of the residual background pressure of order $10^{-8}$ Pa ($10^{-10}$ torr).

The x-ray spectra were recorded with a microcalorimeter which measures the energy of individual photons. As described in more detail in an earlier publication (Tan et al. 2005), the detectors in the microcalorimeter were held at approximately 60 mK using a dual-stage adiabatic demagnetization refrigerator. The lines were identified by calibrating the photon energy scale with reference lines emitted from a solid target of selected materials (including carbon, oxygen, fluorine, copper and aluminum) excited by photons produced by electron impact on a separate target. This two-step process produced clean calibration spectra free of bremsstrahlung background. The observed target was moved in and out of the path between the microcalorimeter and the trapped ions without breaking the vacuum. The calibration elements were separated from both the EBIT and the microcalorimeter by windows, and thus cannot contribute impurity lines to the observed Fe spectrum. Further details and a schematic of the calibration layout can be found in a separate publication (Silver et al. 2010).

In order to get a qualitative overview of the data, we present in figure 1 a composite spectrum obtained by summing all of the data taken at the electron beam energies listed above. The range of photon energies fit in the following analysis is indicated by the extent of the residuals shown. The microcalorimeter itself is sensitive to a much wider range of energies and has essentially a flat amplitude response (not including window transmission) over the limited range shown. Below, we analyze the individual data files separately to precisely determine the energy dependence of the line ratios. The composite spectrum simulates crudely what might have been observed from a plasma with a much wider electron energy distribution than that of our EBIT (350 eV rather than 45 eV); it has a relatively high signal-to-noise ratio, at the expense of electron energy resolution. The spectral line full-width-at-half-maximum is 9 eV, the same as that shown in figure 3 of (Beiersdorfer, et al. 2002). The spectrum is not adjusted for window transmission.

The small peak near the center of figure 1 indicates a very low level of oxygen in the trap (the line is so weak it only appears clearly in the composite spectrum of all the data summed together). It is important to confirm that the oxygen level is low since there is also an oxygen line that lies between the C and D lines that could compromise the results, as we discuss in detail in section 5 below. In the spectra shown in figure 2 and figure 3 of a paper from the LLNL group (Beiersdorfer, et al. 2002), the ratio of the oxygen line to the C line is approximately four times greater than it is in our figure 1.

We also note what appears to be a constant low level offset across the central region of the spectrum shown in figure 1. This is actually a quasi-continuum of real photons (at photon energies above 1.3 keV, only occasional single photon counts can be seen on a zero-noise baseline). The fit area under the very weak E line is approximately equal to the area under the baseline in this region, so we refrain from attempting to determine accurate results for the E line intensity in the present work.

## 3. DATA ANALYSIS

By fitting the spectra from the individual data files rather than summing the data files first, we can determine the energy dependence of the line ratios. The raw data files were thus analyzed individually by least squares fitting the iron lines to a 12 parameter function (5 gaussian profiles of variable center and intensity but one width, and a constant background offset). The G and H lines are known to be so close in energy that they are entirely unresolved in our measurement, and hence fit to a single line. Prior to deciding on the function to fit, we tested the assumption of equal widths by fitting several data sets to lines of variable width, and found all of the widths to be equal to within the statistical uncertainties. We also tested the possibility that the fit to line C was being significantly influenced by the tail of lines at higher energies by truncating the high energy side of the data for this line at various fractions of its peak value, and found no statistically significant changes in the fit line intensity. The results reported below were for the case of data extending to photon energies high enough to cover approximately 95% of the C line intensity. The line ratios reported are the weighted averages of the results of the fits to the individual data sets, with the uncertainties propagated assuming a normal statistical distribution (Bevington 1969). The various systematic corrections discussed below were applied to the weighted averages, and the uncertainties in the corrections propagated to the final results.

### 3.1 Beam Energy

The voltage applied to the EBIT drift tubes creates an electric field which accelerates the electrons to the nominal beam energy. The additional repulsive electric field from the other electrons passing through the trap reduces the actual energy of the incoming electrons by an amount called the space charge shift. This shift can be easily calculated in the limit that the trap contains only electrons and no ions. Between 0-100% of this shift can be cancelled by the neutralizing charge of the trapped ions, however. This contributes to a relatively large uncertainty in the absolute value of the actual electron beam energy. The maximum ("empty trap") shift, S, is calculated at a radius of 33 μm from the trap axis, consistent with the effective electron beam radius measured in a similar trap (Marrs et al. 1995). We assume that all of the electrons are contained inside this radius, and obtain the shift:

$$S = 5.0 \frac{I}{\sqrt{E}} \quad \text{(electron volts)} \qquad (1)$$

where I is the electron beam current in mA, E is the electron beam energy in keV, and the prefactor is a numerical value proportional to the natural logarithm of the inner radius of the center drift tube (5 mm, in our case). We write $S$ as a positive quantity, but note that it enters with a minus sign to always shift the actual energy downwards from the nominal value given by the voltage applied to the drift tubes.

Following the Guidelines for Evaluating and Expressing the Uncertainty of NIST Measurement Results (Taylor & Kuyatt 1994), we adjust our nominal energies, $E_o$ assuming a triangular probability distribution (recommended for the case of 100% probability that the shift has magnitude between 0 and S, but lacking specific knowledge of the exact value in that range other than that the values near the center of the distribution are more likely than the values at the extremes):

$$E = (E_o - \frac{1}{2}S) \pm \frac{1}{2\sqrt{6}}S \qquad (2)$$

The uncertainties specified in the equation above and throughout this paper are at the 68% (1 σ) confidence level. Note that this uncertainty refers to our lack of knowledge of the exact position of the peak of the electron energy distribution function, and not the actual spread in that distribution. The latter is discussed in the following paragraph, while the former corresponds to the error bars shown in the plots below. The beam energy adjustments (*S*/2) range from 33 eV to 55 eV.

The adjusted beam energy, E, represents the peak in a narrow distribution of actual energies. The effective full width at half maximum (FWHM) of this distribution was measured to be 45 eV in a previous experiment (McLaughlin et al. 1996). In that measurement, we scanned a 2.3 keV, 56 mA, electron beam over a set of narrow RR lines and observed the variation in line intensity. As shown below, our present data is consistent with a 45 eV electron energy width as well.

### 3.2 Adjustment for Window Transmission

The windows in our present experiment differ from those used previously. In (Laming, et al. 2000), there were three windows in the microcalorimeter (totaling 240(1) nm of polyimide and 330(2) nm of aluminum) and one additional window separating the EBIT from the microcalorimeter (500 nm of polypropylene). As reported earlier (Silver 2003), the discrepancy between the two experimental groups for the line ratios shown in figure 4 of (Beiersdorfer, et al. 2002) is reduced by approximately one of our experimental error bars after a correction is applied to our data in order to properly account for the EBIT window.

In the present experiment, there are 5 windows in the microcalorimeter (totaling 677(3) nm of polyimide and 266.4(13) nm of aluminum), and one additional window separating the EBIT (97.0(5) nm of polyimide). The total window transmission in the present experiment was approximately 0.3 to 0.4 in the photon energy range of the lines listed in Table 1. The actual value of the transmission of the microcalorimeter windows was monitored during the course of the experiment, and the absolute value was determined as detailed in a separate publication (Silver et al. 2010). In brief, the procedure involved monitoring the intensity of photon energy calibration lines (described above) simultaneously with the microcalorimeter and a SiLi detector whose absolute intensity response was previously calibrated independently. The measured microcalorimeter window transmission was found to be stable, but lower than the transmission calculated using the nominal thicknesses of the windows. The results are shown in figure 2. Had we not measured the actual window transmission but instead relied on the calculated transmission as we did in our earlier work, our present *s*/*C* ratios would have been approximately 7% lower, whereas our *C*/*D* ratios would have been essentially unchanged.

### 3.3 Adjustment for Polarization and Angular Distribution of Radiation

Our microcalorimeter is insensitive to photon polarization, and therefore the largest source of uncertainty in previous measurements using crystal spectrometers (Brown et al., 1998) is entirely absent in the present data (see Beiersdorfer et al. 2002 for a discussion of the magnitude of the polarization uncertainties in their crystal spectrometer measurements). In the comparison to theory, however, it is still necessary to consider the angular distribution of the emitted photons, since the electron beam provides a quantization axis and all EBIT experiments to date have been carried out by observing perpendicular to the electron beam. The Belfast EBIT (Currell et al. 2005) is designed to allow observations at the "magic angle" of 54.7° in order to yield results which are theoretically identical to the case in which observations are averaged over all angles, but results are not yet available from that EBIT.

As described in our original paper (Laming, et al. 2000), the angular distribution of radiation was included as part of our theoretical predictions, rather than applied as an adjustment to our measurements. The LLNL group has taken the opposite view, however, and applied an adjustment to their data to convert it into what they predict they *would have observed* if their experiment had been done at the magic angle (or averaged over an ensemble of randomly oriented angles). This adjustment is one reason why the often-reproduced Figure 4 in (Beiersdorfer, et al. 2002) indicates a larger discrepancy between the two experimental groups than appropriate. The NIST data shown in that plot are the actual values measured at the observation angle of 90 degrees, while the LLNL data have been adjusted (in the direction away from our results) by approximately 13%. Correcting the plot for this disparity reduces the previous discrepancy between the NIST and LLNL results by approximately one combined standard uncertainty at 900 eV. To compare to the LLNL results in the present paper, we have divided our experimental results by the following factors which arise from applying the angular emission factors discussed in the theory section below to the relevant line ratios: 0.88(2) for *s/C*, and 1.001(4) for *C/D*. These correction factors are consistent with the data from all three models shown in table II.

4 THEORY

We present here new results from calculations by the Auburn-Rollins group, the SAO group, and the NRL group, following on their earlier work (Loch, et al. 2006), (Chen 2008b), and (Laming, et al. 2000), respectively. All of these calculations include the effect of radiative cascades, but to varying levels of completeness, as disussed below.

The NRL calculations include all excitations up to principal quantum number n=5 and the strongest excitations for n=6. The initial population by electron impact is split up among the m-sublevels, and radiative cascades computed by m-sublevel (rather than by j-level). Collisional depopulation is also included, but has very little effect at the assumed density of $1 \times 10^{10}$ cm$^{-3}$. This density, as well as that in the present experiment (approximately $1 \times 10^{12}$ cm$^{-3}$ to $2 \times 10^{12}$ cm$^{-3}$, for 33 μm beam radius) are both in the low-density limit. Density dependent effects are expected at higher densities, where collisonal deexcitation begins to compete with radiative decay for the $2s^2 2p^5 3s$ $^3P_0$ level.

In the Auburn-Rollins calculations, collisional excitation from the ground state and radiative cascades are included to up to n=5. Collisional depopulation/redistribution of the excited states is not included since the radiative lifetimes are much shorter than any collisional-depopulation effects (coronal approximation). This calculation predicts that collisional redistribution of the excited states begins to become important above $10^{15}$ cm$^{-3}$.

In the SAO calculations, levels up to n=10 have been included in the radiative cascades. The electron density is assumed to be $10^{13}$ cm$^{-3}$, and the effect of collisional depletion on C/D is less than 1% in the range $10^9$ cm$^{-3}$ to $10^{14}$ cm$^{-3}$ (Chen 2008b).

## 4.1 Einstein A-Coefficients

For the present paper, the Auburn-Rollins group has updated their previously published results (Loch, et al. 2006). The changes include new values of the Einstein A-coefficients for the C and D lines, and the extension of the calculation to higher electron beam energies and to a variety of assumed gaussian electron beam energy widths. The use of revised A-coefficients was inspired by discrepancies of up to 10% between the previously used values (Loch, et al. 2006) and the values calculated in more recent work by one of us (Chen 2007, 2008b). The problem was traced to a very slowly converging configuration-interaction, as described in section 2 of (Griffin et al. 2008). The new values of A-coefficients are 2.69x10$^{13}$ s$^{-1}$ for the C line and 6.30x10$^{12}$ s$^{-1}$ for the D line. The new A-coefficients did not change the C/D ratio because the upper states in the C and D lines are populated largely by direct excitation from the ground state (little effect from cascades from higher levels). The new results are presented in section 4.3 below.

## 4.2 Angular Distribution of Radiation

Some of the previous theoretical work has presented line ratios that are averaged over all angles of observation relative to the quantization axis (defined by the velocity vector of the incident electron). This is appropriate for a non-aligned thermal plasma, but for comparison to the actual observations from an EBIT experiment in which the electrons are uniaxial and the observations are made at 90 degrees to the axis, the calculation must be multiplied by the factor,

$$M = \frac{3}{3-P} \quad (3)$$

which is given in terms of the polarization parameter,

$$P = \frac{I_{para} - I_{perp}}{I_{para} + I_{perp}} \quad (4)$$

The subscripts on the intensities refer to the orientation of the photon electric field vector (not propagation direction) parallel to and perpendicular to the electron beam direction respectively, when the photons are emitted in a direction perpendicular to the electron beam direction.

For future reference (for adjusting other calculations, or experiments) we tabulate our predictions for this M-factor, under three different assumptions. Predictions for the M-factor, following our earlier work (Laming, et al. 2000) based on distorted wave (DW) cross sections (Bhatia & Doschek 1992), are listed in table II and plotted in figure 3. When we substitute EIE cross sections calculated in the R-matrix approximation (Mohan, Sharma, & Eissner 1997) in place of the DW cross sections, the results, shown in the column of table II labeled RM, agree to 1% or better. Both of these calculations predict that M deviates from unity by less than 1.4% for each of the three *s* lines but by as much as 16% for the two *d* lines calculated. We have also calculated M using the Flexible Atomic Code (FAC) code developed by a member of the LLNL team (Gu 2008), and we find agreement with the DW values for each of the *d* lines to within 0.7%, and for each of the *s* lines to within 3.5% (also shown in table II and figure 3). The calculations were all done under the assumption that the electrons are traveling in a straight line (in an actual EBIT experiment, the electrons can spiral around the magnetic field lines if not launched into the field coaxially). We refer to a very recent paper by some of us for more results on calculation of the polarization of the emitted radiation (Chen et al. 2009).

### 4.3 Line Ratio Predictions

The last three columns of table II gives the line intensities predicted by the NRL group in units that can be used to estimate the absolute emission rates in experiments for which the effective electron density is known; they are angularly averaged, and so should be adjusted by the M-factors before comparing to observations at 90 degrees. Note that while the associated angular correction factors (M) shown in the same table agree fairly well between the various NRL calculations, the absolute values of the predicted intensities vary by as much as 80% and the intensity ratios vary by as much as 40%.

The predictions of the Auburn-Rollins group and the SAO group are shown in figure 4 and figure 5 below. Earlier predictions for the *C/D* ratio, reviewed in (Beiersdorfer 2003), have ranged from 2.9 to 4.7.

### 5. EXPERIMENTAL RESULTS AND DISCUSSION

Our results, after adjustment for the systematic effects described above, are shown in comparison to some other results in figures 6 and 7. Our line ratios, as well as the new theoretical results, are all calculated in terms of number of photons detected and are therefore 2% lower (*C/D*) and 14% higher (*s/C*) than they would have been if calculated in terms of the classical line intensity defined by the energy detected (Cowan 1981). Agreement of our results with the closest theory, under the assumption of 45 eV to 60 eV beam energy width, is typically better than 10%, and agreement with the experimental results from LLNL (Beiersdorfer, et al. 2002; Brown 2008; Brown, et al. 2006; Brown, et al. 1998) is typically within the combined error bars. Exceptions and systematic deviations are discussed below.

For the *C/D* ratio, the two calculations shown in figure 4 typically agree to approximately 10%, with the results of Chen being systematically lower than the results of the Auburn-

Rollins group. In figure 6, the results of our experiment are in agreement with the LLNL EBIT results that have the smallest vertical error bars (Brown, et al. 1998), but our results are systematically lower for all energies. The lower of the two LLNL data points at 1150 eV was taken in a separate experiment using a Bragg crystal that was different from that used for the other three points (Brown, et al. 1998). The overall results of the most accurate experiments favor the predictions of Chen over those of the Auburn-Rollins group, particularly if the horizontal error bar in the lowest energy points are taken into account.

Also shown in figure 6 are some other LLNL results (Brown 2008) taken with a different technique and which have larger error bars. For these data (shown as small squares), the electron beam was not held stationary, but swept through a wide range of energies and the data segregated into energy bins of width indicated by the horizontal error bars. At higher energies (up to 2.6 keV), there are 5 additional data points for C/D given by (Brown 2008) that have been interpreted as a systematic variation from theory. These points gradually fall to 2.4 then rise back up as high as 3.4, but most of these points overlap with the predictions of one of us (GXC) to within the error bars (two are lower by 1.3 and 2.5 experimental error bars), so additional work at high energies may be needed to determine if this deviation is significant.

Not shown in figure 6 are the values reported by LLNL in 2001(Brown, et al. 2001a; Brown, et al. 2001b). One of these concluded that there was no energy dependence in the range studied (1000 eV to 3800 eV) and averaged all 22 values measured (ranging from 2.3 to 4.0) to obtain a smaller uncertainty on the reported ratio *C/D*=3.04(19). The other of these two papers (Brown, et al. 2001a), tabulates a *C/D* value which is identical in magnitude and uncertainty to the average of their two earlier values shown in our figure 6 at 1150 eV, although the later paper (Brown, et al. 2001a) specifies the beam energy to be 10 eV lower and the beam energy uncertainty to be 30% larger.

Our lowest energy point in figure 6 deserves some specific remarks as it corresponds to a considerably lower line ratio than the other points. We note that this point lies in a region for which the line ratio is very sensitive to the energy. If the data point is shifted both to lower energy and higher C/D ratio, by an amount equal to our horizontal and vertical error bars, it nearly touches the current predictions of the Auburn-Rollins group. We await an extension of the predictions of Chen to lower energies to see if this theory will continue to predict somewhat lower ratios.

For the *s/C* ratio (figure 7), calculations by Chen are still in progress, so we compare only to those of the Auburn-Rollins group assuming 45 eV beam energy width (FWHM). We note that for this line ratio, unlike the case for *C/D* discussed above, both the NIST data and the most accurate LLNL data were taken with the same type of detector (microcalorimeters) with similar resolution. For clarity, we do not show all of the LLNL grating and crystal spectroscopy results, which are in agreement with their other results shown but have error bars that are a factor of 1.5 to 1.75 larger. We divide the discussion into three sections: (1) low electron energy (E<870 eV) in which the theory predicts a sharp rise, (2) medium electron energy (870 eV to 1200 eV) in which potentially problematic oxygen lines may have affected earlier data and (3) high electron energy (E>1200 eV) in which new channels due to inner shell ionization of Fe XVI exist (Sampson & Zhang 1987) but are not included in the present calculations.

In the low energy range, our data are in agreement with the predictions for the 45 eV beam energy width previously measured in our EBIT (McLaughlin, et al. 1996). While the LLNL data point in this energy range (Beiersdorfer, et al. 2002) disagrees with the 45 eV curve by nearly 12 times the quoted experimental uncertainty, it would seem more likely that their energy for this point is somewhat off. Their paper states that the uncertainty in energy is negligible, however, because the effective space charge shift is measured rather than calculated. Although they did not specify the energy error bars explicitly, they cited an earlier paper which describes the technique used to measure the energy (Gu et al. 2001) and this technique was later quoted to an accuracy of 5 eV (Brown, et al. 2006). This uncertainty is seven to eleven times smaller than the shift needed to bring the low energy LLNL point into agreement with the 45 eV prediction. We thus consider an alternative explanation which seems more likely, namely that their data were taken with a beam energy width much narrower than 45 eV. Their paper does not specify the width, but in describing the beam temperature they cite their earlier papers in which the width was measured to be 50 eV (Beiersdorfer et al. 1992) and 35 eV (Gu, et al. 2001). In a later paper (Brown, et al. 2006), however, they measured the width to be 20 eV. Because the width depends on electron current and other experimental parameters, it is reasonable to expect variations between experiments. Note that their beam energy is only 4 eV above threshold for excitation of the *C* line, so if their beam energy width were significantly greater than 8 eV (or perhaps 15 eV after reasonable uncertainties are taken into account), a significant fraction of their electrons would be below threshold to excite the *C* line, and so their *s/C* ratio would be much higher.

Assuming a relatively narrow beam energy width in the LLNL experiment (Beiersdorfer, et al. 2002), as discussed above, we are able to obtain consistency between both the LLNL data and the present NIST data by using the results predicted for 15 eV width shown by the dotted curve in figure 7 (for clarity, we only show the dotted curve for energies below 850 eV; at higher energies, the difference between the two curves, as shown in figure 5, is approximately equal to the typical size of the experimental error bars). We suggest that figure 7 explains the otherwise inconsistent results shown from the LLNL and NIST groups as being largely due to different widths of the electron beam energy distributions. While the rise in *s/C* and the reduction in *C/D* for our lowest energy points could be largely the result of a reduction in C due to part of the electron energy distribution falling below threshold, our point at 857 eV shows no such reduction in *C/D*, suggesting that the rise in *s/C* is due primarily to excitation resonances.

Although the magnitude of the increase in s/C shown in our data in going from 902 eV to 857 eV suggests that resonance enhancement is indeed significant, as predicted, smaller error bars are necessary to conclude this definitively. Detailed plots of various calculations published previously (Chen 2007; Chen & Pradhan 2002; Loch, et al. 2006) allow one to see the relative contributions of the resonant structure to the nonresonant structure and suggest that a significant part of the predicted rise in the *s/C* ratio is a result of enhancement of the *s* lines in the numerator, rather than reduction in the *C* line in the denominator. This is consistent with the behavior of our C/D ratios shown in figure 6, as well as the general conclusion that resonances affect dipole allowed transitions less strongly than they do forbidden transitions (Chen & Pradhan 2002). The relatively low value of the LLNL point at 830 eV is consistent with the predicted lack of significant resonance enhancement at this particular energy, together with the assumption of an energy width of approximately 15 eV or less to avoid sampling the large density of resonances that occur above and below this particular energy.

In the medium energy range, the NIST data agree with theory to within their error bars, while the LLNL data are systematically low for most of the region, by somewhat more than their error bars. One effect that can alter the *s/C* ratio is the presence of oxygen or lower charge states of iron. Above 740 eV beam energy, for example, hydrogen-like O VIII can be produced, leading to 1s-4p and 1s-5p lines that straddle the Fe XVII C line by -9 eV and +10.6 eV. The weak line in the center of figure 1 is the 1s-3p line of the same O VIII Rydberg series, and can thus be used to gauge the presence of oxygen in this charge state.

We have modeled the effect that oxygen could have on the Fe XVII line ratios by generating synthetic spectra for various relative concentrations of iron and oxygen and fitting them assuming only iron is present. The results suggest that the *C/D* ratio from the experiment of (Brown, et al. 1998) is unaffected by the oxygen level indicated there, but that the *s/C* ratio from the experiment of (Beiersdorfer, et al. 2002) could be low by 9% or more. For our present experiment, the oxygen content changes our results for the *C/D* and *s/C* ratios by much less than our uncertainty due to statistical noise. To our knowledge, none of the previous results have been adjusted for oxygen content. We find that constraining the linewidths to one common fit value, as we have done, helps reduce the sensitivity of the ratios to oxygen concentration.

In the high energy range, where there is no predicted resonant structure, both the LLNL data and the NIST data suggest that the theory is systematically low by 10-15%. Note that the predictions of the Auburn-Rollins group differed from the predictions of Chen by about this much for the case of the *C/D* ratio (figure 6). In this range, additional channels not included in the calculations will open up but they are expected to contribute to shifts only on the order of a few percent in the direction of our data (Loch 2010). These channels include (1) resonances (e.g. the $2s2p^6$ n=5 up to n=8 resonances extend up to approximately 1335 eV) and (2) cascades from n>5. In addition, inner-shell ionization of Fe XVI, producing excited states of Fe XVII, could contribute if there is a significant population of the lower charge state ion in the trap, and if the beam energy is high enough to remove electrons from the n=2 shell. Previous tabulations of inner-shell channels (Sampson & Zhang 1987) indicate that they should begin around 1200 eV, at about the upper limit of the horizontal error bar of our highest energy point. Improved sets of calculations of the line ratios would be useful in this region.

The energies of the potentially problematic inner shell satellite lines observed from lower charge states by (Brown, et al. 2001a), are indicated along the top margin of figure 1 as circles (Fe XVI) and square (Fe XV). There is no evidence for the presence of these lines in the residuals to our fit, consistent with the finding of (Brown, et al. 2001a) that these lines are not present when MEVVA ion injection is used to load the trap. In contrast, when the trap is loaded with gas injection, it was found (Brown, et al. 2001a) that these lines can make the *C/D* ratio appear artificially low (<2). Simulations of the charge state distribution indicate that the equilibrium fraction of Fe XVI (and lower charge states) in our trap is a few percent or less for background pressures on the order of $10^{-8}$ Pa ($10^{-10}$ torr), typical for the MEVVA ion injection that we use. This level of Fe XVI would have an insignificant affect on our line ratios, in comparison to the magnitude of our error bars. A higher fraction of Fe XVI exists briefly during the ionization buildup in the trap, but since our observation time per injection (several seconds) is approximately one hundred times longer than the time to reach collisional equilibrium, our results will not be significantly affected by the short-time transient. As assessed previously (Gillaspy et al. 2004), gating out the small fraction of observation time that occurs during and shortly after the injection of ions into the trap, as routinely done by the LLNL group, is thus not a significant difference between the results from the two groups, unless the equilibrium time is long, for example, due to the presence of a high background pressure.

We believe the experimental results presented here represent the most stringent test of the calculated line ratios for Fe XVII for the following combined set of reasons: (1) they are taken with a microcalorimeter, which avoids relatively large polarization adjustments (some of which depend on EBIT operating conditions), (2) they do not involve gas injection, which has been shown to be an EBIT operating condition that can lead to inaccurate line ratios in Fe XVII, (3) they remove uncertainties in our previous data concerning the window transmission, (4) they are taken under conditions of much lower oxygen background than some previous work, and (5) they explicitly specify uncertainties in the beam energy.

Before concluding, we turn to the claim by (Brown, et al. 2006) that the absolute values of the calculated EIE cross sections for the lines *C* and *D* are in error by 25% to 50%, despite the relatively good agreement in the ratios of these lines. This claim was based on their measurement of the ratio of EIE lines to the RR lines, the latter of which they deemed to be accurate to 5% and then used to infer values of the former to an overall accuracy of 10-20% (after taking into account five other sources of experimental uncertainty). Their work has been criticized, however, for underestimating the uncertainty in the calculated RR rates (Chen 2007). In a subsequent publication (Chen 2008a), a relativistic close-coupling calculation of the RR cross sections for the $3d_{3/2}$ and $3d_{5/2}$ states, based on the advanced Breit-Pauli R-matrix method, indicated that indeed the normalization to the 3d lines used by (Brown, et al. 2006) should be adjusted upwards by 24%. If the new values for the RR cross sections are convolved with an electron energy distribution of FWHM 20 eV (equal to that specified in the experiment), and then used in place of those calculated by (Brown, et al. 2006), then the measurements of (Brown, et al. 2006) can be reinterpreted as providing a 3C EIE cross section of 11.1(23) x $10^{-20}$ cm$^2$ at 964 eV electron impact energy. The uncertainty on this result includes (in quadrature) a 7.5% uncertainty for the new RR cross section, as given by (Chen 2008a). This revised

3C EIE cross section is in agreement with the value predicted by the converged Dirac R-matrix calculation: 10.7(5) x $10^{-20}$ cm$^2$ (Chen 2007).

In retrospect, it appears that the original paper by(Scofield 1976), which detailed the theory used to calculate the RR cross sections by (Brown, et al. 2006), was remarkably prescient in noting that while the total predicted cross sections agreed with experiments to 5%, the subshell cross sections were off by approximately 30%. A more recent comprehensive comparison of the predictions of Scofield's theory to experiments for a wide range of elements and energies (Saloman, Hubbell, & Scofield 1988) provides many examples in which the predictions deviate from experiment by much more than 5%. Furthermore, we note that all of the available comparisons of the predicted RR cross sections based on the theory of (Scofield 1976) given by (Saloman, et al. 1988) are for the case of neutral atoms. For the case of ions as highly charged as Fe XVII, the predicted subshell photoionization cross sections have remained virtually untested in the laboratory.

In order to have an independent calculation of the RR subshell cross sections for all of the states of Fe XVII used by (Brown, et al. 2006), we have examined the results of (Trzhaskovskaya, Nikulin, & Clark 2008). After converting these results to RR cross sections and applying the correction factor 3/(3-P) using the polarization factors given by (Brown, et al. 2006) to adjust for the fact that the observations were limited to 90°, we find that the results of (Trzhaskovskaya, et al. 2008) are lower than the values given by (Brown, et al. 2006) by 35% for the s-state, 6% for the sum of the p-states, and 7% for the sum of the d-states. The same theory, however, agrees with the theory of (Scofield 1976) for the case of neutral atoms to better than 3%, even for the case of subshell cross sections. For the 3s state of Fe XVII, we can also compare to the results of (Donnelly et al. 1999), which are in better agreement with (Trzhaskovskaya, et al. 2008) than they are with (Brown, et al. 2006). Thus it appears that the uncertainties in the *total* cross sections for *neutral* atoms are not good predictors of the uncertainties for *subshell* cross sections of *highly charged ions*. This underscores the need for more work on measuring photoionization and/or RR cross sections of highly charged ions in order to more precisely test the variety of available calculations.

We can also obtain a new value for the 3D EIE cross section from the data of (Brown, et al. 2006) by using their directly measured C/D ratio and the new value for the 3C EIE cross section given above. If we assume that their reported uncertainty in the C/D ratio is independent of their uncertainty in the EIE/RR ratio, then their 3D EIE cross section becomes 3.75(86) x $10^{-20}$ cm$^2$ at 964 eV electron impact energy. If we assume that the uncertainties in their C/D and EIE/RR ratio are fully correlated such that the C/D uncertainty cancels out, then the result is 3.75(77) x $10^{-20}$ cm$^2$. The actual uncertainty will be somewhere in between these two limits. The new EIE 3D cross section agrees with the value predicted by the converged Dirac R-matrix calculation: 3.91(20) x $10^{-20}$ cm$^2$ (Chen 2007).

We conclude that the current state-of-the-art calculations have now converged to the measured values (at the level of 5%-20%, depending on the sophistication of the calculation and the energy range) for the strongest lines in Fe XVII excited by electron impact, and can therefore be used for precise plasma diagnostics as long as adequate attention is given to assumptions pertaining to the (a) peak and width of the electron energy distribution, (b) directional distribution of the electron velocities relative to the observation angle, (c) absorption of material intermediate between the emitter and the detector (including vacuum windows, in the case of laboratory measurements), and (d) possible line overlaps from other charge states of Fe or other species (such as oxygen). We hope that the removal of large (of order 100%) discrepancies discussed in the literature previously makes these lines now ready for widespread use as a strong and reliable plasma diagnostic.

Acknowledgements:

We thank Stuart Loch and Connor Ballance for their calculations (done in collaboration with Mitch Pindzola and Don Griffin), and Don Landis, Norm Madden, Jeff Beeman, Eugene Haller, Gerry Austin, and David Caldwell for contributing to the development of the microcalorimeter. This work is supported by NASA grant NNX08AK33G.

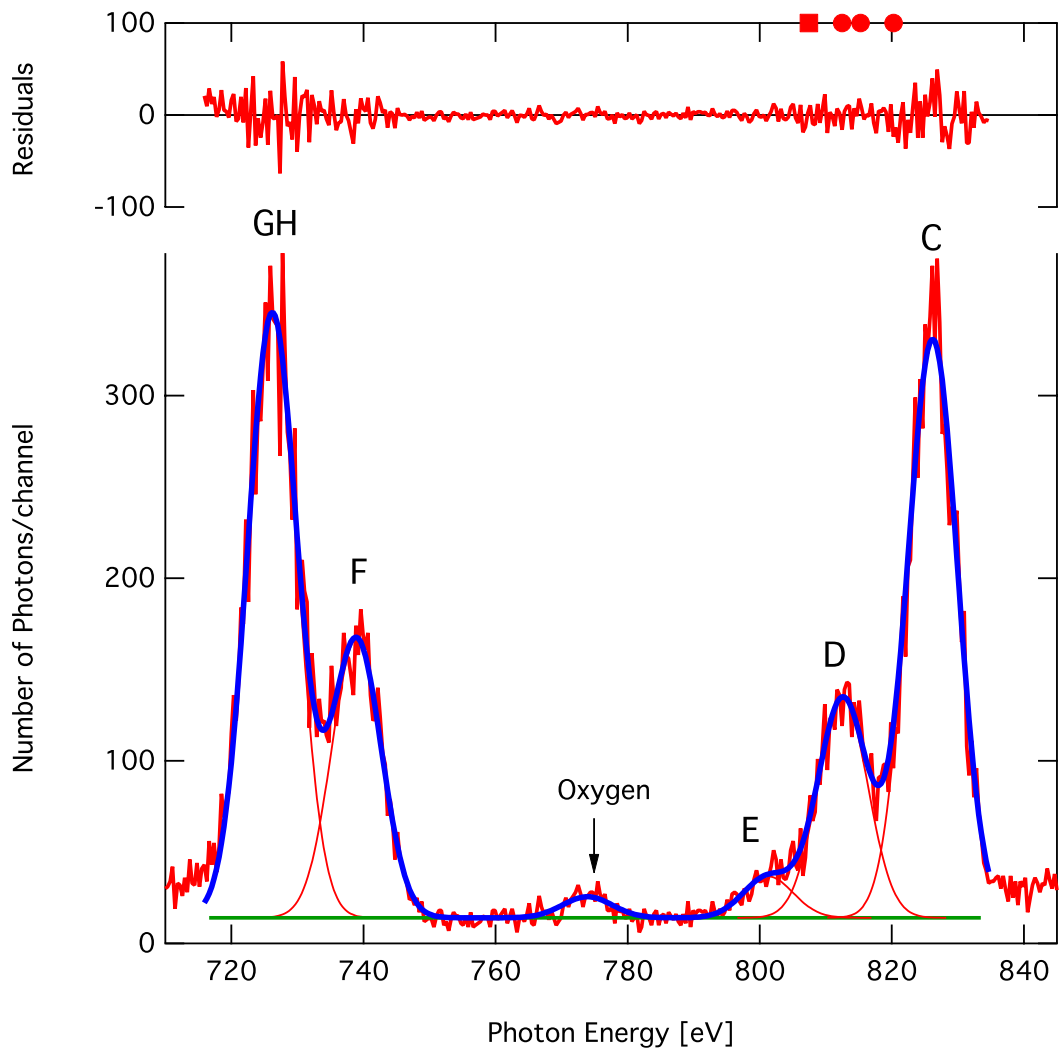

**Figure 1.** Composite spectrum of raw data in the region of the lines shown in Table I. Also shown are five gaussian profiles fit to the iron lines, plus a sixth profile fit to the oxygen line. The widths of the lines are all fixed to one value which is determined by the fit.

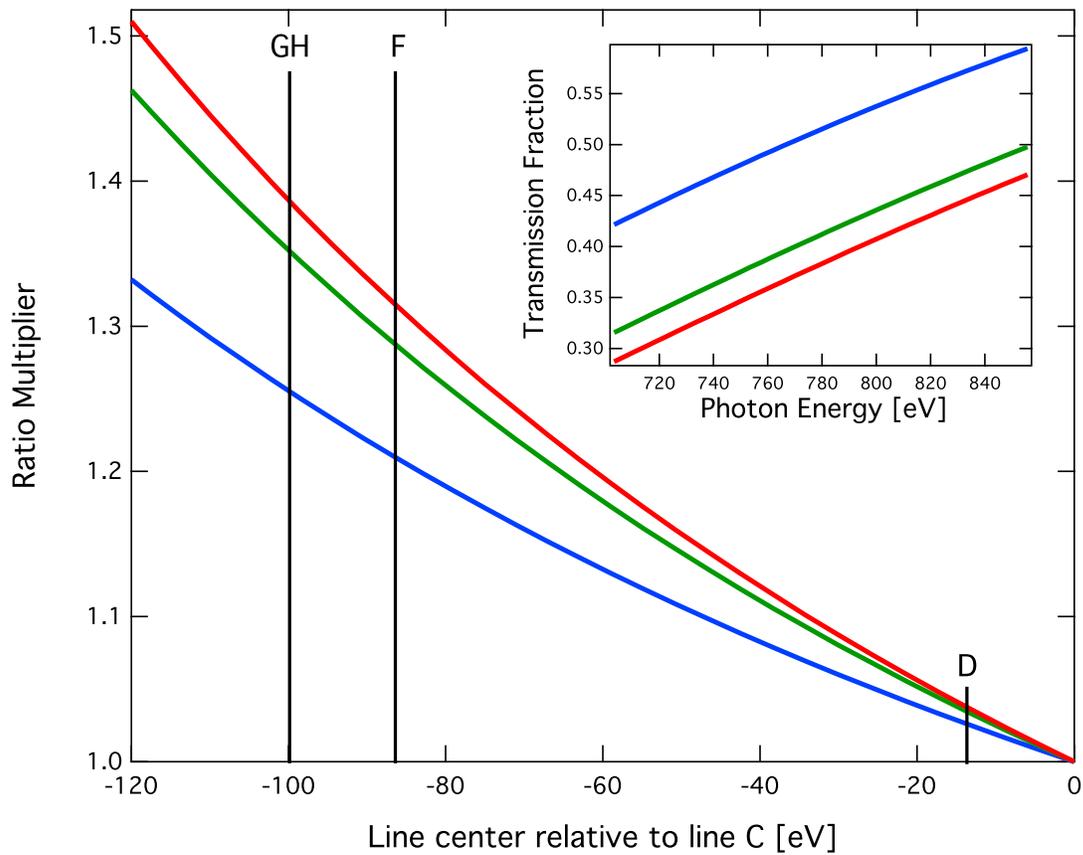

**Figure 2.** Window transmission corrections for line intensities relative to the C line. The upper curve (red) is the total correction applied to the data including the EBIT and microcalorimeter windows. The lower curve (blue) is the calculated value for the microcalorimeter windows and the middle curve (green) is the measured value for the microcalorimeter windows. D, F, and GH mark the positions of the corresponding lines in table I. Inset shows the transmission of the microcalorimeter windows calculated (upper, blue) and measured (middle, green), as well as the total microcalorimeter and EBIT window transmission used in the data analysis (lower, red).

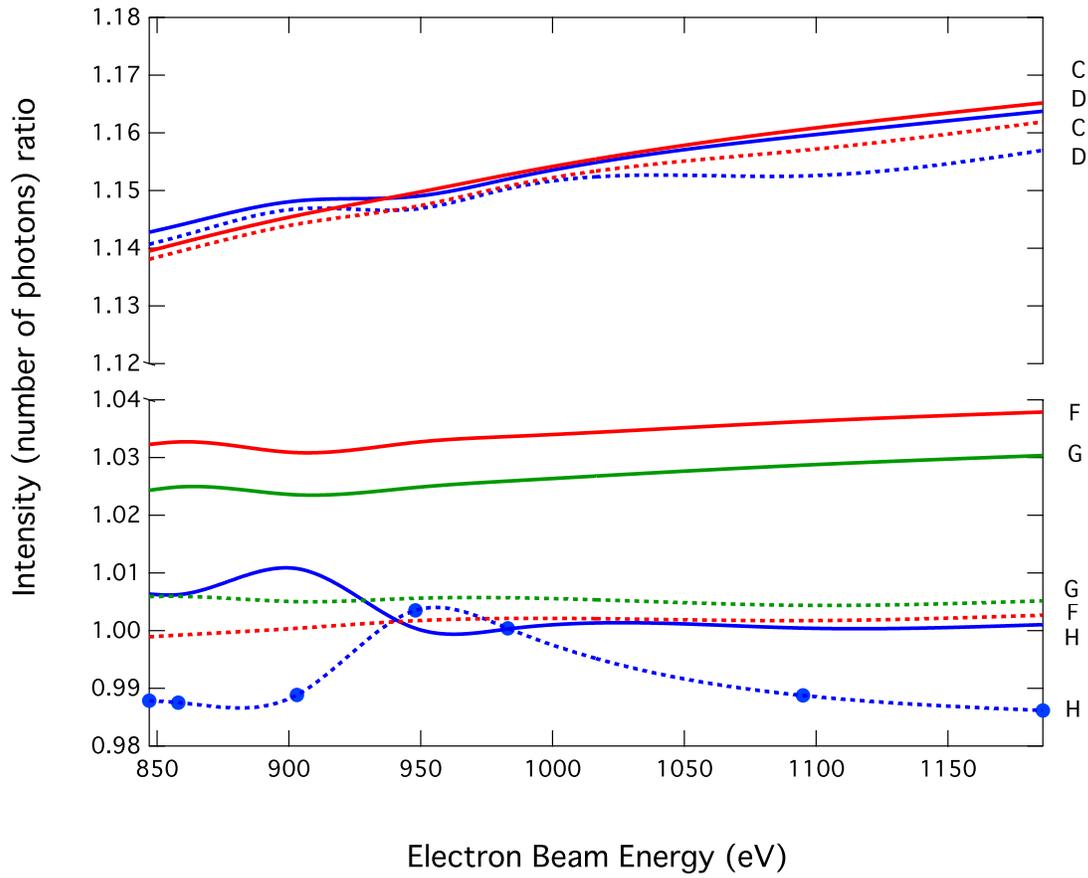

**Figure 3.** Calculated anisotropy multipliers, M (equation 3), as a function of electron beam energy. Curves are cubic spline interpolations of the DW (dotted line) and FAC (solid line) data shown in table II at the energies indicated by the circles on one of the curves.

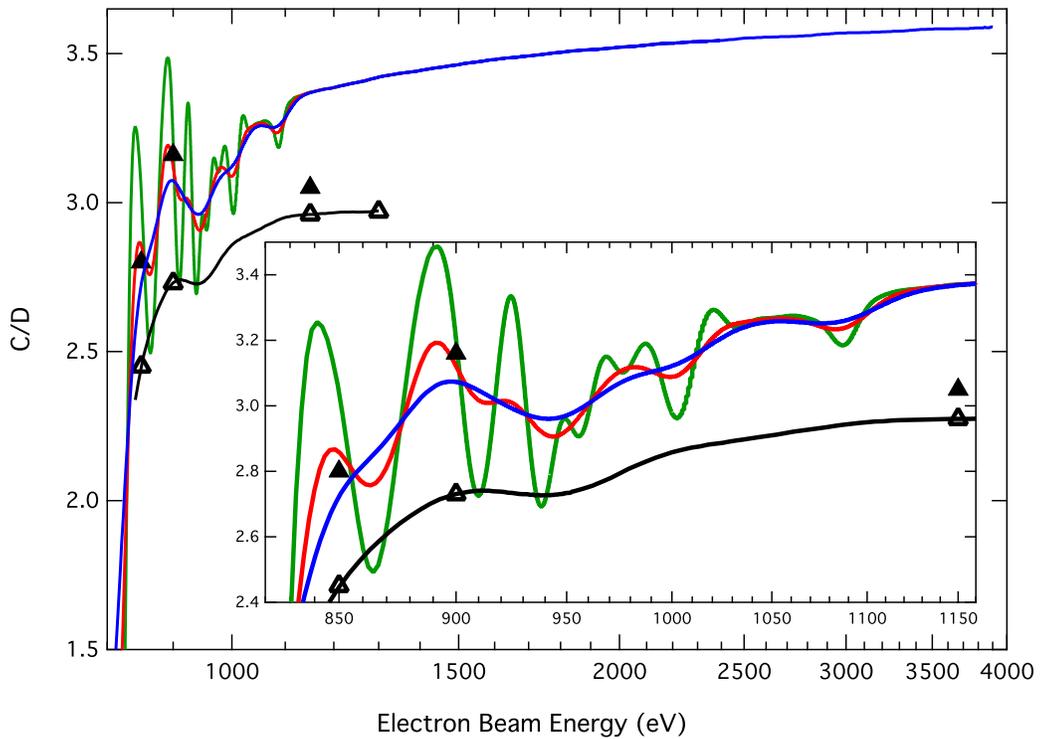

**Figure 4.** Predictions for the *C/D* ratio. Upper curves are the calculations of the Auburn-Rollins group presented here for the first time at high energies and for various beam energy widths (FWHM): 15 eV (most strongly oscillating, green), 30 eV (intermediate oscillations, red), and 45 eV (least strongly oscillating, blue). Lower curve is the present Dirac R-matrix calculation of one of us (GXC) for 60 eV (FWHM) beam energy width; the open triangles are the results of this calculation published earlier at the 4 energies indicated (Chen 2008b). Closed triangles are previous Breit-Pauli calculations for 30 eV beam energy width (Chen & Pradhan 2002). The inset shows a closeup of the curves in the region of oscillation.

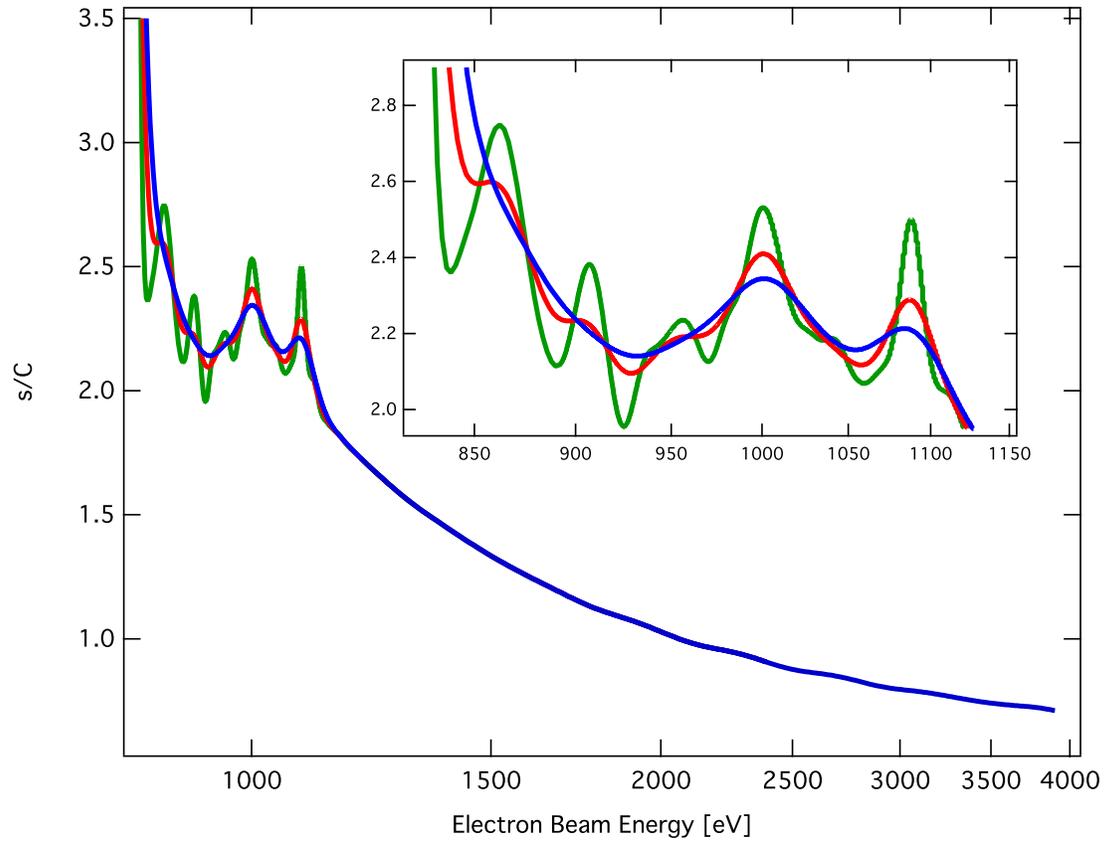

**Figure 5.** Recent theoretical predictions for the *s/C* ratio. Upper curves are previously unpublished results of the Auburn-Rollins group for various beam energy widths (FWHM): 15 eV (most strongly oscillating, green), 30 eV (intermediate oscillations, red), and 45 eV (least strongly oscillating, blue). The inset shows a closeup of the region of oscillation.

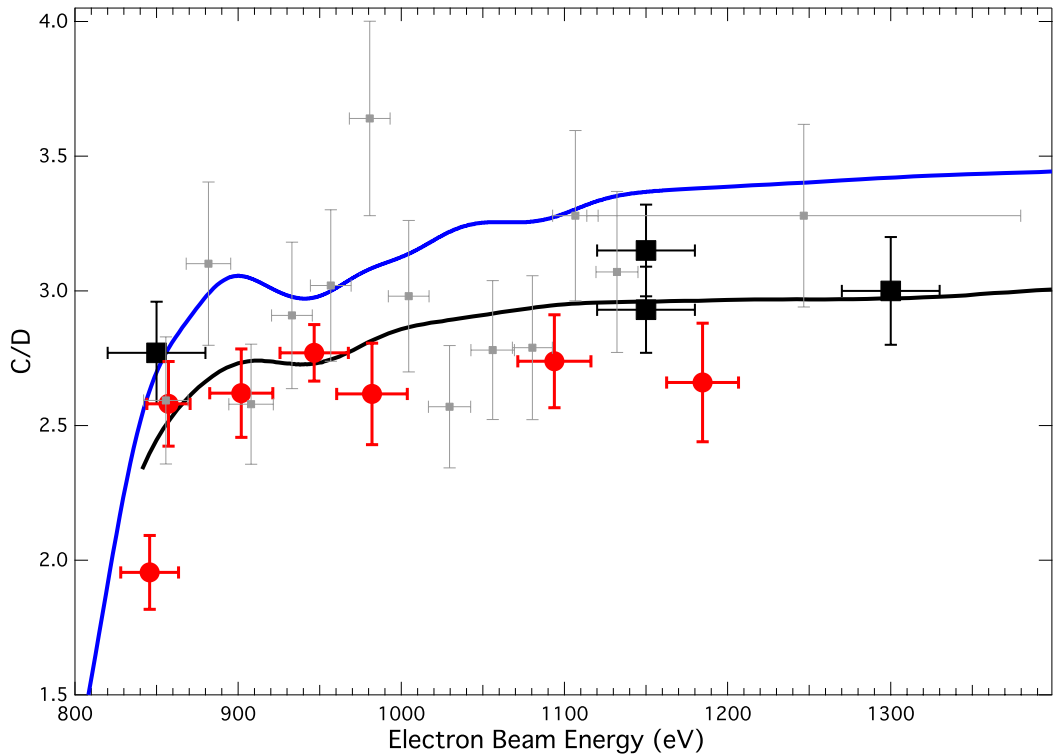

**Figure 6.** Comparison of theory and the most accurate experiments for the *C/D* ratio. The present experimental data are shown as circles (red), and the LLNL data (Brown, et al. 1998) are shown as large squares. Also shown as small squares are some other LLNL data (Brown 2008) with larger error bars. The upper curve (blue) shows the theory from the Auburn-Rollins group following their earlier work (Loch, et al. 2006), convoluted with 45 eV beam energy width (FWHM). The lower curve is the present calculation of one of us (GXC) convoluted with a 60 eV beam energy width (FWHM). The energy error bars for the dots and the large squares represent the uncertainty in the value of the most probable energy, not the beam energy width. For the small squares, the beam energy was swept through the width of the error bars shown.

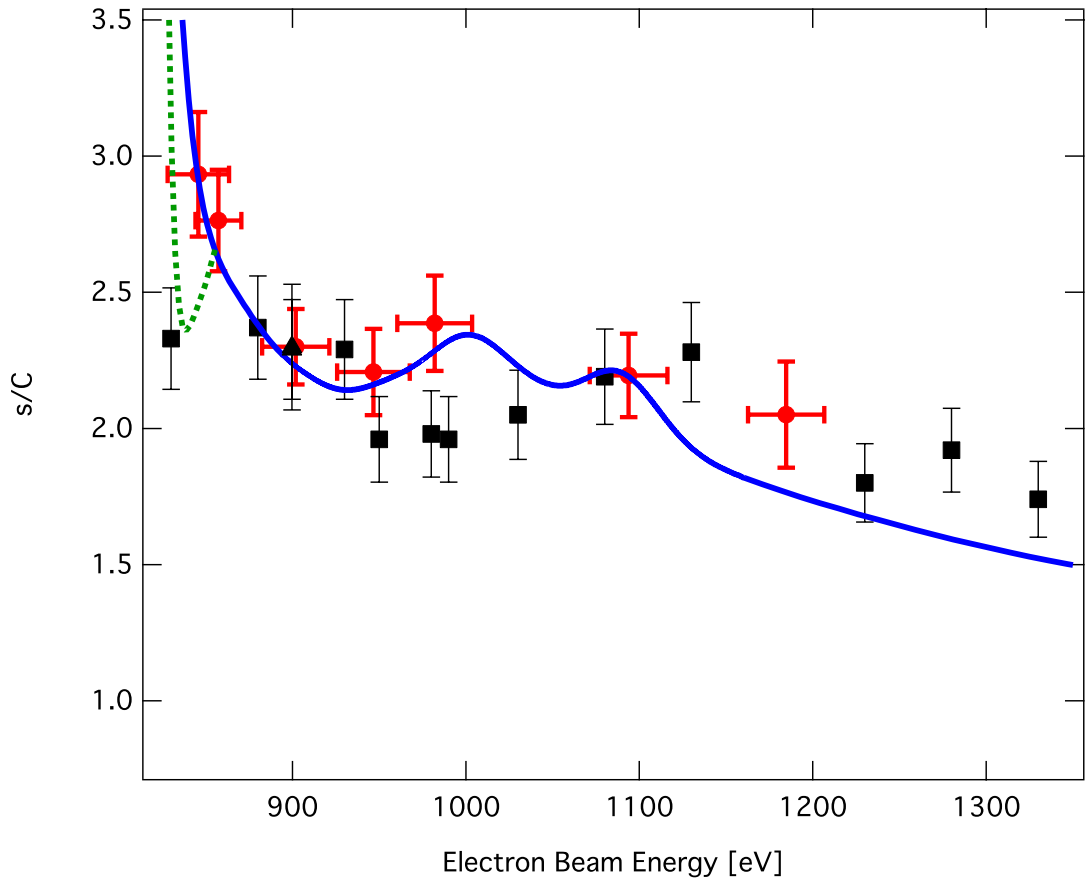

**Figure 7.** Comparison of theory and the most accurate available experiments for the *s/C* ratio. Present experimental results are shown as dots (red) and previous microcalorimeter results from the LLNL group (Beiersdorfer, et al. 2002) are shown as squares. Also shown (triangles) are two LLNL points recorded with crystal and grating spectrometers at the 900 eV energy discussed above (for which there is no LLNL microcalorimeter data available). Predictions of the Auburn-Rollins group, presented here following their earlier work (Loch, et al. 2006), are shown as a solid line (blue) for 45 eV beam energy width (FWHM) and for low energies also as a dotted line (green) for 15 eV beam energy width. The energy error bars (not available for the LLNL results, but specified as "negligible") represent the uncertainty in the value of the most probable energy, not the beam energy width.

**Table I.**
Spectral lines in Ne-like Fe discussed in this paper.

| Group | Line | Quantum Numbers | Wavelength (Å) | Transition Energy (eV) |
|---|---|---|---|---|
| d | C | $2p^6\ ^1S_0 - 2p^5\ 3d\ ^1P_1$ | 15.01 | 826 |
|   | D | $2p^6\ ^1S_0 - 2p^5\ 3d\ ^3D_1$ | 15.26 | 812 |
|   | E | $2p^6\ ^1S_0 - 2p^5\ 3d\ ^3P_1$ | 15.45 | 802 |
| s | F | $2p^6\ ^1S_0 - 2p^5\ 3s\ ^3P_1$ | 16.78 | 739 |
|   | G | $2p^6\ ^1S_0 - 2p^5\ 3s\ ^1P_1$ | 17.05 | 727 |
|   | H | $2p^6\ ^1S_0 - 2p^5\ 3s\ ^3P_2$ | 17.10 | 725 |

**Table II.**
Calculated values for the angular emission correction factor (M) and the line intensity (number of photons emitted per ion, per unit electron density).

| Line | Beam Energy (eV) | M (Equation 3) | | | Line Intensity ($10^{-12}$ s$^{-1}$ cm$^3$) | | |
|---|---|---|---|---|---|---|---|
|   |   | RM | DW | FAC | RM | DW | FAC |
| C | 847 | 1.14 | 1.14 | 1.14 | 202 | 211 | 211 |
|   | 858 | 1.14 | 1.14 | 1.14 | 202 | 211 | 212 |
|   | 903 | 1.14 | 1.14 | 1.15 | 204 | 213 | 217 |
|   | 948 | 1.15 | 1.15 | 1.15 | 209 | 217 | 223 |
|   | 983 | 1.15 | 1.15 | 1.15 | 211 | 218 | 226 |
|   | 1095 | 1.16 | 1.16 | 1.16 | 225 | 228 | 235 |
|   | 1186 | 1.16 | 1.16 | 1.17 | 232 | 232 | 242 |
| D | 847 | 1.14 | 1.14 | 1.14 | 52.4 | 53.7 | 58.3 |
|   | 858 | 1.14 | 1.14 | 1.14 | 52.4 | 53.8 | 58.4 |
|   | 903 | 1.15 | 1.15 | 1.15 | 52.8 | 54.2 | 59.4 |
|   | 948 | 1.15 | 1.15 | 1.15 | 54.8 | 56 | 62.4 |
|   | 983 | 1.15 | 1.15 | 1.15 | 55.3 | 56.3 | 63 |
|   | 1095 | 1.15 | 1.15 | 1.16 | 59.5 | 59.8 | 64.7 |
|   | 1186 | 1.16 | 1.16 | 1.16 | 60.8 | 60.6 | 65.9 |
| F | 847 | 1.01 | 1.00 | 1.03 | 86.2 | 88.8 | 108 |
|   | 858 | 1.01 | 1.00 | 1.03 | 85.7 | 88 | 107 |
|   | 903 | 1.01 | 1.00 | 1.03 | 97.1 | 100 | 122 |
|   | 948 | 1.01 | 1.00 | 1.03 | 103 | 107 | 129 |
|   | 983 | 1.01 | 1.00 | 1.03 | 101 | 105 | 127 |
|   | 1095 | 1.01 | 1.00 | 1.04 | 117 | 124 | 121 |
|   | 1186 | 1.01 | 1.00 | 1.04 | 115 | 123 | 117 |
| G | 847 | 1.01 | 1.01 | 1.02 | 103 | 96.1 | 134 |
|   | 858 | 1.01 | 1.01 | 1.02 | 102 | 95.3 | 133 |

|   | 903  | 1.01 | 1.01 | 1.02 | 120  | 114  | 156 |
|---|------|------|------|------|------|------|-----|
|   | 948  | 1.01 | 1.01 | 1.02 | 127  | 123  | 167 |
|   | 983  | 1.01 | 1.01 | 1.03 | 125  | 121  | 163 |
|   | 1095 | 1.01 | 1.00 | 1.03 | 135  | 137  | 154 |
|   | 1186 | 1.01 | 1.01 | 1.03 | 135  | 140  | 149 |
| H | 847  | 0.99 | 0.99 | 1.01 | 89.1 | 71.4 | 127 |
|   | 858  | 0.99 | 0.99 | 1.01 | 88.1 | 70.7 | 125 |
|   | 903  | 0.99 | 0.99 | 1.01 | 87.4 | 71.6 | 132 |
|   | 948  | 1.00 | 1.00 | 1.00 | 99.2 | 85.7 | 148 |
|   | 983  | 1.00 | 1.00 | 1.00 | 95.8 | 84.3 | 142 |
|   | 1095 | 0.99 | 0.99 | 1.00 | 101  | 96.5 | 128 |
|   | 1186 | 0.99 | 0.99 | 1.00 | 100  | 100  | 118 |